\documentclass[useAMS,usenatbib]{mn2e}

\usepackage{subfigure,epsfig,amsmath,amssymb,lastpage} 
\title[2-D homography-based correction of positional errors in widefield 
MRT images] {Two-dimensional homography-based correction of positional 
errors in widefield MRT images}

\author[A. Nayak, S. Daiboo and N. Udaya Shankar]
{Arvind Nayak,$^{1}$\thanks{E-mail: arvind@rri.res.in}
Soobash Daiboo$^{1,2}$ and N. Udaya Shankar$^{1}$\\
$^{1}$Raman Research Institute, C.V. Raman Avenue, 
Sadashivanagar, Bangalore 560 080, India\\
$^{2}$Physics Department, University of Mauritius, Reduit, Mauritius}

\begin{document}
\date{Accepted 2010 June 09. Received 2010 June 04; in original form 
2010 February 02}
\pagerange{\pageref{firstpage}--\pageref{lastpage}} \pubyear{2010}

\maketitle
\label{firstpage}

\begin{abstract}
A steradian of the southern sky has been imaged at 151.5~MHz using the
Mauritius Radio Telescope (MRT). These images show systematics in
positional errors of sources when compared to source positions in the
Molonglo Reference Catalogue (MRC). We have applied two-dimensional
homography to correct for systematic positional errors in the image domain
and thereby avoid re-processing the visibility data. Positions of bright
(above 15-$\sigma$) point sources, common to MRT catalogue and MRC, are
used to set up an over-determined system to solve for the homography
matrix. After correction the errors are found to be within 10\% of the
beamwidth for these bright sources and the systematics are eliminated from
the images. This technique will be of relevance to the new generation
radio telescopes where, owing to huge data rates, only images after a
certain integration would be recorded as opposed to raw visibilities.
It is also interesting to note how our investigations cued to possible
errors in the array geometry. The analysis of positional errors of sources
showed that MRT images are stretched in declination by $\sim 1$ part in 1000.
This translates to a compression of the baseline scale in the visibility
domain. The array geometry was re-estimated using the astrometry principle.
The estimates show an error of $\sim 1$~mm/m, which results in an error of
about half a wavelength at 150~MHz for a 1~km north-south baseline. The
estimates also indicate that the east-west arm is inclined by an angle of
$\sim 40\arcsec$ to the true east-west direction.
\end{abstract}


\begin{figure*}
\centering
\subfigure[]{
\epsfig{figure=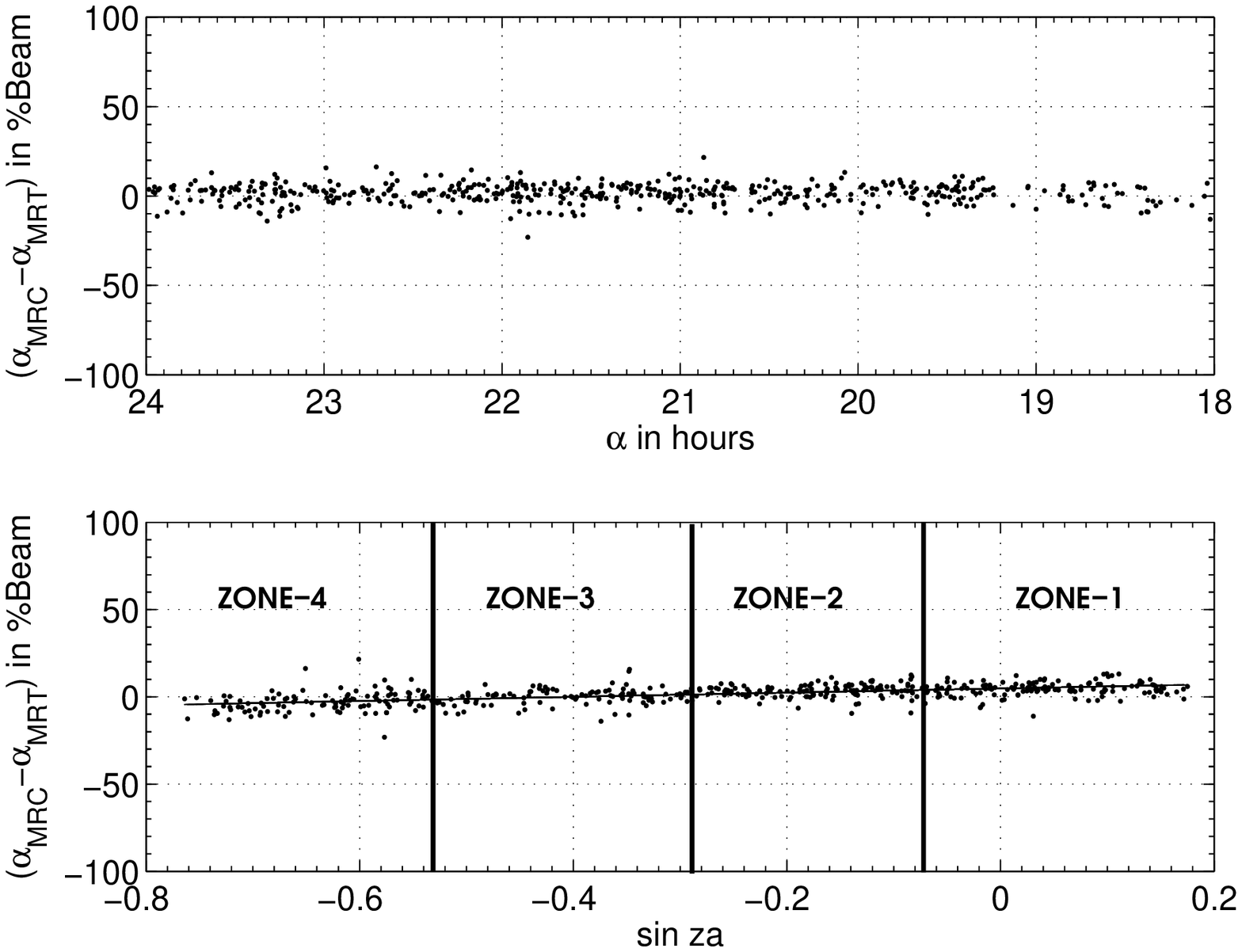,width=0.44\linewidth}}
\subfigure[]{
\epsfig{figure=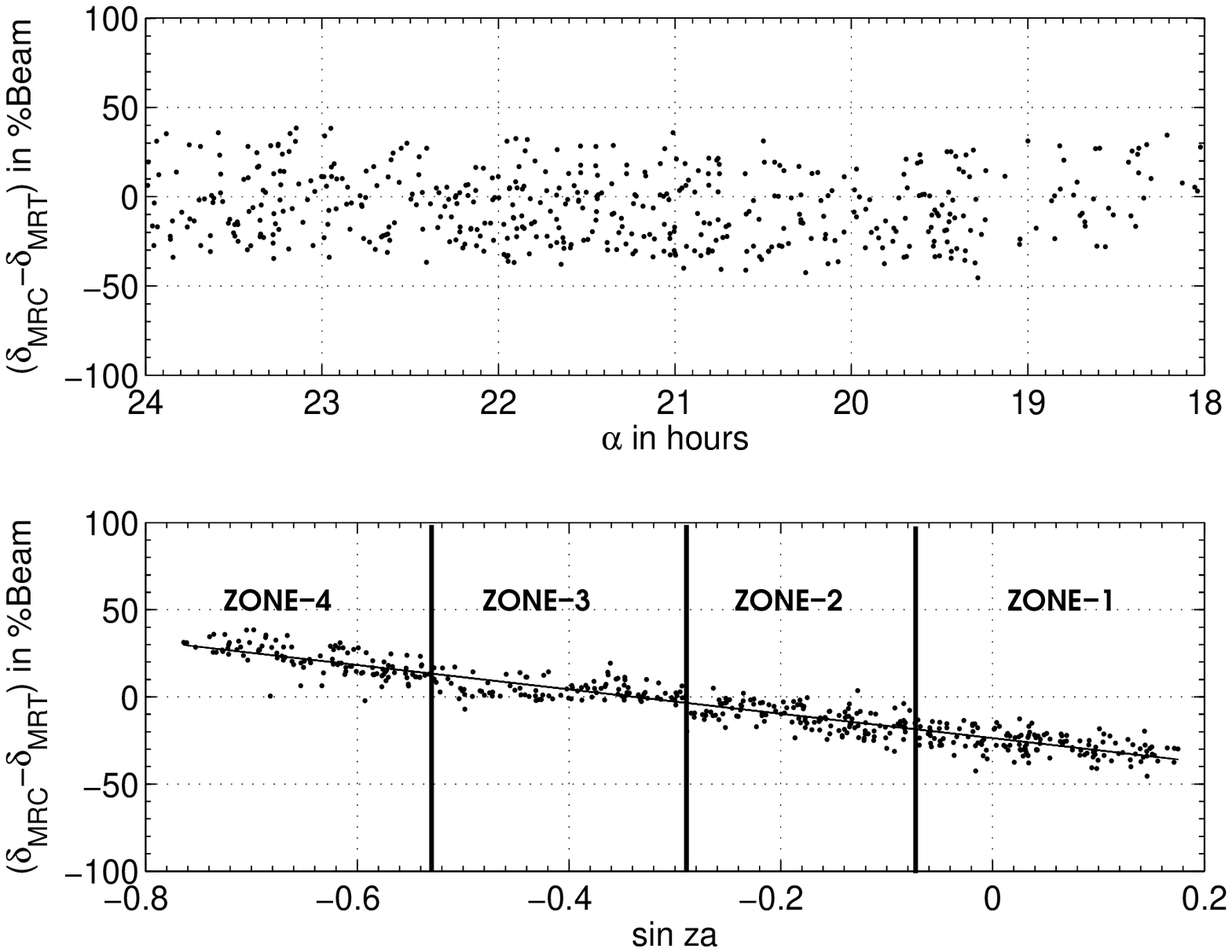,width=0.44\linewidth}}
\subfigure[]{
\epsfig{figure=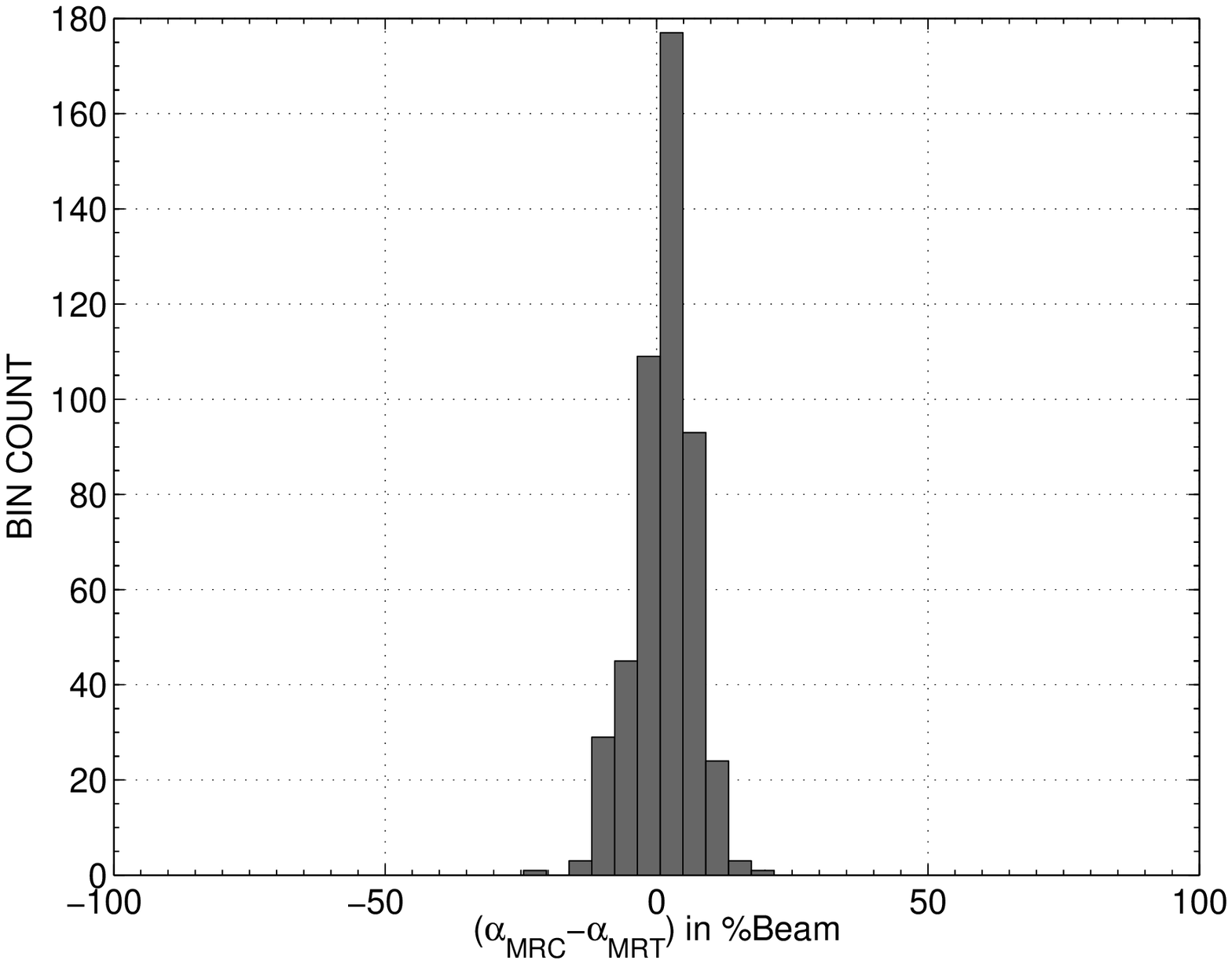,width=0.44\linewidth}}
\subfigure[]{
\epsfig{figure=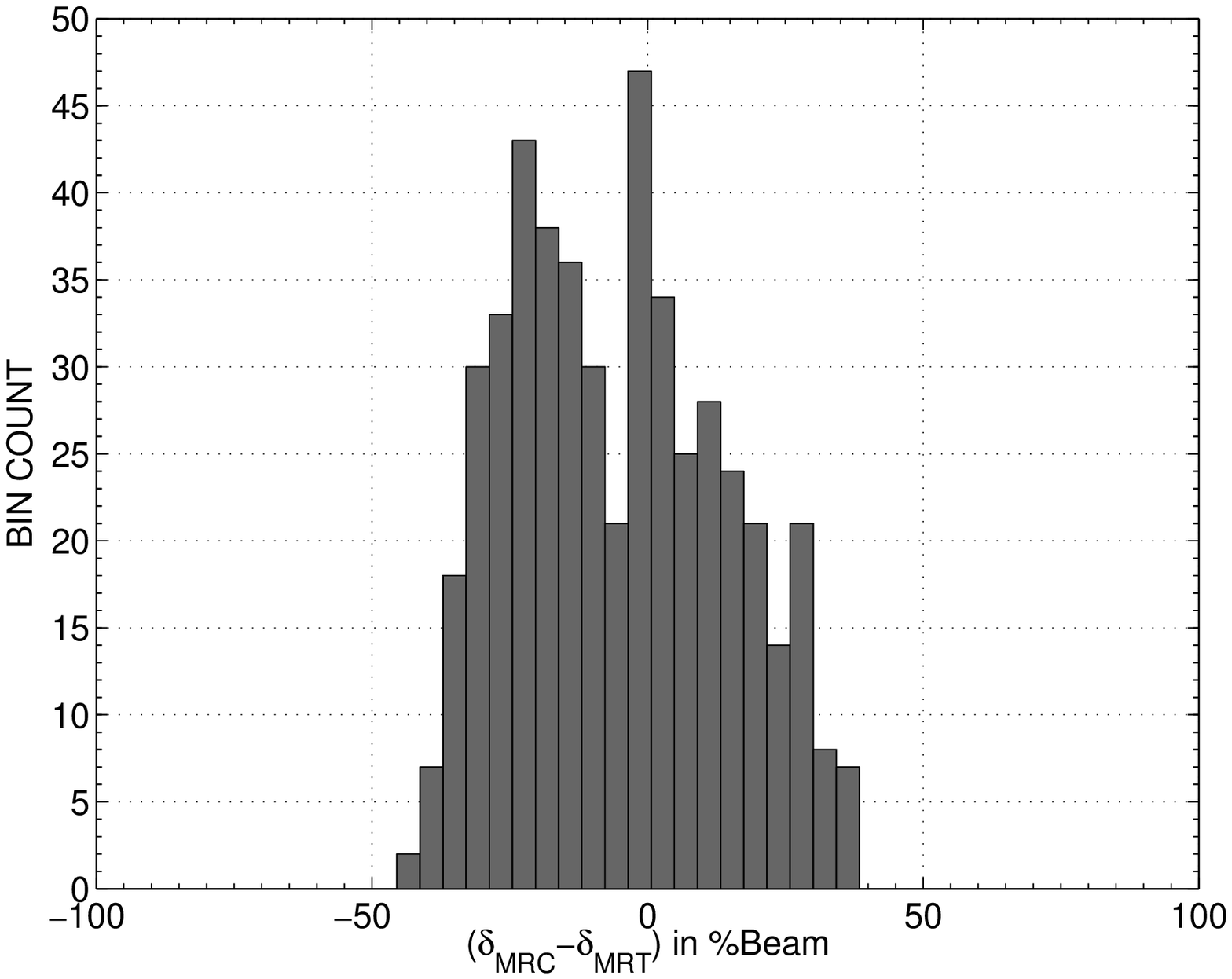,width=0.44\linewidth}}
\caption{ \small 
Positional error analysis of $\sim 400$ sources (above 15-$\sigma$) common 
to MRT catalogue and MRC. For visualisation, the errors
are shown in percentages of MRT beamwidths.
{\bf (a)} The {\it first row} subplot shows errors in $\alpha$
against $\alpha$; no systematics are observed. 
The {\it second row} subplot shows 
errors in $\alpha$ against $\sin za$; errors show a linear gradient as a 
function of $\sin za$.
{\bf (b)} The {\it first row} subplot shows errors in $\delta$
against $\alpha$; no systematics are observed. The {\it second row} subplot shows 
errors in $\delta$ against $\sin za$; errors show a linear gradient as a function 
of $\sin za$. The {\it second row} subplots in (a) and (b) also indicate declination 
(or equivalent $\sin za$) ranges of the four {\it zones} imaged with different delay 
settings.
{\bf (c)} and {\bf (d)} show histograms of errors in $\alpha$ and $\delta$,
respectively. The histogram of 
errors in $\delta$ shows a broader spread compared to errors in $\alpha$.\normalsize }
\label{f:sourcecomparison}
\end{figure*}

\section{Introduction}
\label{s:introduction}
The Mauritius Radio Telescope (MRT)~\citep{ias:golap95a,apss:uday02} is
a Fourier synthesis, T-shaped non-coplanar array operating at 151.5~MHz.
The telescope was built to fill the gap in the availability of deep
sky surveys at low radio frequencies in the southern hemisphere. The aim
of the survey with MRT is
to contribute to the database of southern sky sources in the declination
($\delta$) range $-70^\circ$ to $-10^\circ$, covering the entire right ascension
($\alpha$), with a synthesised beam of
$4\arcmin\times4\arcmin.6\sec za$ and an expected point source
sensitivity (1-$\sigma$) of $\sim110$~mJy~beam$^{-1}$.
The {\it zenith angle} $(za)$ is given by $(\delta - \phi)$, 
where, $\phi$ $\left(\approx -20.14^\circ\right)$ is the latitude of MRT.
MRT has been designed to be the southern-sky 
equivalent of the Cambridge 6C survey at 151.5~MHz~\citep{mnras:baldwin85}.

The next generation radio telescopes, like the LOw Frequency ARray (LOFAR)
and the Murchison Widefield Array (MWA), that are being built are low 
frequency arrays; clearly indicating a renewed interest in metre-wavelength 
astronomy. The key astrophysical science drivers include acceleration, turbulence 
and propagation in the galactic interstellar medium, exploring the high 
redshift universe and transient phenomenon, as well as searching for the 
redshifted signature of neutral hydrogen from the cosmologically important 
epoch of reionisation (EoR). The surveys made using such arrays will 
provide critical information about foregrounds which will also provide a 
useful database for both extragalactic and galactic sources. MRT survey at 
151.5~MHz is a step in that direction and, in addition, will provide the 
crucial sky model for calibration.

Imaging at MRT is presently done only on the meridian to minimise the problems
of non-coplanarity. A two-dimensional (2-D) image in $\alpha$-$\sin za$ 
coordinates is formed by stacking one-dimensional
(1-D) images on the meridian at different sidereal times.
Images of $\sim$ a steradian $(18^{\mbox{\small h}}\leq\alpha\leq24^{\mbox{\small h}},-70^{\circ}\leq\delta\leq-10^{\circ})$ of the southern sky, with an rms noise in images
of $\sim 300$~mJy~beam$^{-1}$ (1-$\sigma$), were produced 
by~\cite{ursiga:pandey05}.
A suite of programs developed in-house was used to reduce $\sim 5000$~hours 
of the survey data
(a quarter of the total $\sim 20,000$ 
hours observed over a span of $\sim 5$ years). 
The deconvolved images and a source catalogue of 
$\sim 2,800$ sources were published by~\cite{thesis:pandey06}.

Systematics in positional errors were found when the positions of sources
common to MRT catalogue and the Molonglo Reference Catalogue (MRC) 
\citep{mnras:large81} were compared. 
\cite{thesis:pandey06} treated the systematics in errors in $\alpha$ and $\sin za$ 
independently. By estimating two separate 1-D least-squares fits for 
errors in $\alpha$ and $\sin za$ the systematics were corrected only in the 
source catalogue. However, errors remained in the images which impede 
usefulness of MRT images for multi-wavelength analysis of sources.
In addition, the source of errors was not investigated. At MRT, the 
visibility data is processed through several complex stages of data 
reduction specific to the array, especially, arising due to its 
non-coplanarity~\citep{apss:uday02}. It was therefore decided to correct 
for errors in the image domain and avoid re-processing the visibility 
data.

This paper describes the application of 2-D homography, 
a technique ubiquitous in the computer vision and graphics community, to 
correct the errors in the image domain. Homography is used to estimate a
transformation matrix (which includes rotation, translation and
non-isotropic scaling) that accounts for positional errors in the
linearly gridded 2-D images.
In our view, this technique will be of relevance to the new generation 
radio telescopes where, owing to huge data rates, only images after a 
certain integration would be recorded as opposed to raw 
visibilities~\citep{ieee:lonsdale09}. This paper also describes our 
investigations tracing the positional errors to errors in the array 
geometry used for imaging. Our hypothesis on the array geometry, its 
subsequent confirmation endorsed by re-estimation of the array geometry 
and its effect on the images are also described.

The rest of the paper is organised as follows. Section~\ref{s:poserror} 
compares positions of sources common to MRT catalogue and MRC. The 2-D 
homography estimation is briefly described in Section~\ref{s:homography}. 
Section~\ref{s:scheme} presents the correction 
scheme and typical results. The re-estimation of MRT array 
geometry is described in Section~\ref{s:arraygeometry}. Finally, we 
summarise and present our conclusions in Section~\ref{s:conclusions}.

\section{Positional errors}
\label{s:poserror}
The positions of sources common to MRT catalogue and MRC were compared.
We used MRC because of its overlap with MRT survey,
its proximity in frequency compared to other reliable catalogues
available and, comparable resolution $(2\arcmin.62\times2\arcmin.86\sec(\delta+35^\circ.5))$.
Moreover, for sources of listed flux density~$\geq 1.00$~Jy 
(at 408~MHz) the catalogue is reported to be substantially
complete and, the reliability is reported to be 99.9\%~\citep{mnras:large81}.
For our further discussions, errors in MRC source positions are 
considered random, without any systematics.

About 400 bright sources common to the two catalogues and with flux density
at 151.5~MHz greater than 5~Jy ($> 15$-$\sigma$) were identified and their 
positions were compared.
The sources were labelled as common if they lie within $4\arcmin$ of each other. 
Since MRC has a source density of $\sim0.5$~source~deg$^{-2}$, 
the chances of considering two unrelated sources as common are extremely low.
A flux threshold of 15-$\sigma$ ensures a source population 
abundant to reliably estimate homography (explained in next section).

The positional errors in $\alpha$ and $\delta$ show
no systematics as a function of $\alpha$ (refer {\it first rows} of 
Fig.~\ref{f:sourcecomparison}a~and~\ref{f:sourcecomparison}b). 
For visualisation, the errors are shown in percentages of MRT beamwidths.
The errors in $\alpha$ and $\delta$ show a linear gradient as a function of 
$\sin za$. The errors in $\alpha$, plotted against $\sin za$, reach $\sim \pm10\%$ of the 
MRT beamwidth (refer {\it second row} of Fig.~\ref{f:sourcecomparison}a).
Whereas, the errors in $\delta $, plotted against $\sin za$, are
significant and reach $\sim \pm50\%$ of MRT beamwidth.
(refer {\it second row} of Fig.~\ref{f:sourcecomparison}b). 
Histograms in Fig.~\ref{f:sourcecomparison}c~and~Fig.~\ref{f:sourcecomparison}d 
show the distribution of errors in $\alpha$ and $\delta$, respectively. The histogram of 
errors in $\delta$ shows a broader spread compared to errors in $\alpha$.

Re-imaging, to correct for errors in the images, would involve 
re-reducing the $\sim 5,000$ hours of observed data.
Owing to the complexity involved it was decided to correct for the positional
errors in the images, thus avoiding re-processing.
The 2-D homography estimation technique was employed for correcting positional 
errors in images and is discussed in detail in the following section.

\begin{figure*}
\centering
\epsfig{figure=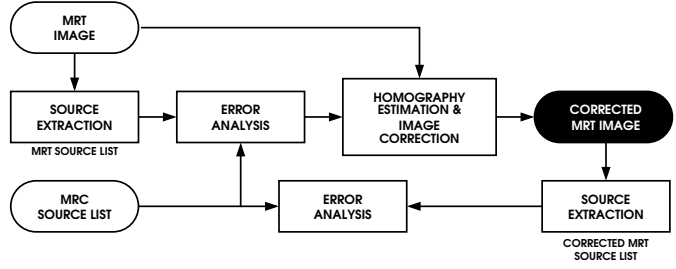,width=0.5\linewidth}
\caption{ \small Block schematic of the correction scheme. Rectangular
boxes represent processes; rounded boxes, data or results. \normalsize }
\label{f:scheme}
\end{figure*}

\section{2-D homography}
\label{s:homography}

The 2-D planar homography is a non-singular linear relationship between points
on planes. Given two sets of $K$ corresponding image points in projective
coordinates, $({\mathbf p}^{}_{k} \mbox{ and } {\mathbf p}'_{k}) \in \mathbb{P}^{2}$, homography maps
${\mathbf p}^{}_{k}$ to the corresponding ${\mathbf p}'_{k}$ \citep{book:hartley}.
Where, $k=1,\ldots,K$.
The homography sought here is a non-singular $3\times 3$ matrix $\tt H$ such
that:
\begin{equation}
\left[ \begin{array}{c}
x'_{k}\\y'_{k}\\1 
\end{array}\right]=
\left[ \begin{array}{ccc}
h^{}_{11}&h^{}_{12}&h^{}_{13}\\
h^{}_{21}&h^{}_{22}&h^{}_{23}\\
h^{}_{31}&h^{}_{32}&h^{}_{33}
\end{array}\right]
\left[ \begin{array}{c}
x^{}_{k}\\y^{}_{k}\\1\end{array}\right].
\label{eq:inhomog}
\end{equation}
\noindent Where, $(x^{}_{k},y^{}_{k})$ and $(x'_{k},y'_{k})$ represent $(\alpha,\sin za)$ of $K$ corresponding
MRT and MRC sources, respectively. 

In Equation~\ref{eq:inhomog}, $\left(x^{}_{k},y^{}_{k},1\right)$ and $\left(x'_{k},y'_{k},1\right)$ are 
referred to as the {\it homogeneous coordinates} and are always represented one dimension higher than 
the dimension of the problem space. This is a commonly used representation in computer graphics. 
The simple reason is that with a $2\times2$ matrix one can only {\it rotate} a set of 2-D
points around the origin and {\it scale} them towards or away from the origin. A $2\times2$
matrix is incapable of {\it translating} a set of 2-D points. The homogeneous coordinates 
allow one to express a translation as a multiplication. A single $3\times3$ matrix,
with homogeneous coordinates, can account for rotation, scaling and translation of 2-D
coordinates.
For example, from Equation~\ref{eq:inhomog}, $x'_{k} = h^{}_{11}x^{}_{k} + h^{}_{12}y^{}_{k} + h^{}_{13}$.
Notice, $h^{}_{13}$ (representing translation in 
$\alpha$-dimension) is simply being added to the normal dot product $(h^{}_{11}x^{}_{k} + h^{}_{12}y^{}_{k})$
that together represents rotation and scaling.
In homogeneous coordinates, the 2-D problem space is a plane hovering in the
third dimension at a unit distance.

A general homography matrix, for projective transformation, has 8 degrees-of-freedom (DOF).
For our system, both errors in $\alpha$ and $\delta$ have only $\sin za$-dependency. Therefore, a less general, 
2-D affine transformation is sufficient. A 2-D affine 
transformation (two rotations, two translations and two scalings)
requires 6-DOF~\citep{book:hartley}, therefore in $\tt H$, $h^{}_{31}$ and $h^{}_{32}$ are zero.
Since each 2-D point provides two independent equations, a minimum of 
3 point correspondences are necessary to constrain $\tt H$ in the affine space. 
A set of $K$ such equation pairs, contributed 
by $K$ point correspondences, form an over-determined linear system: 
\begin{equation}
\nonumber
{\tt A}{\mathbf h}={\mathbf b},\,\,\mbox{where},
\end{equation}
\begin{equation}
\nonumber
{\tt A}=
\left[ \begin{array}{cccccccc}
x^{}_{1} & y^{}_{1} & 1 & 0 & 0 & 0 & -x^{}_{1}x_{1}' & -x_{1}'y^{}_{1}\\
0 & 0 & 0 & x^{}_{1} & y^{}_{1} & 1 & -x^{}_{1}y_{1}' & -y^{}_{1}y_{1}'\\
\vdots &&&& \vdots&&&\vdots\\
x^{}_{K} & y^{}_{K} & 1 & 0 & 0 & 0 & -x^{}_{K}x'_{K} & -x'_{K}y^{}_{K}\\
0 & 0 & 0 & x^{}_{K} & y^{}_{K} & 1 & -x^{}_{K}y'_{K} & -y^{}_{K}y'_{K}\\
\end{array}\right]\mbox{, }
\end{equation}
\begin{equation}
\nonumber
{\mathbf h}=
\left[\begin{array}{ccc}
h^{}_{11},
h^{}_{12},
h^{}_{13},
h^{}_{21},
h^{}_{22},
h^{}_{23},
h^{}_{31},
h^{}_{32}\end{array}\right]^{T}\mbox{and, }
\end{equation}
\begin{equation}
\label{e:system}
{\mathbf b}=
\left[\begin{array}{c}
x'_{1},y'_{1},\ldots,
x'_{K},y'_{K}
\end{array}\right]^{T}.
\end{equation}
\noindent In Equation~\ref{e:system}, $T$ represents transpose of a matrix. This 
system can be solved by least squares-based estimators.

At this stage it is useful to consider the effect of using ($\alpha, \sin za$)-coordinates
to represent the brightness distribution on the celestial sphere. 
Ideally, it is the directional cosines $(l,m,n)$, with respect to the coordinates 
of the array, which represent the spherical coordinates in the sky. Therefore, the 
image coordinates in which homography should in principle be estimated 
are $(l,m)$. However, at MRT, for 1-D imaging on the meridian:
\begin{equation}
m = \sin za. 
\end{equation}
\noindent 
Therefore, $\sin za$ is a natural choice for one of the coordinates and is indeed
used in the present case.
On the meridian, the directional cosine $l$ is 
zero. For small errors, $\Delta l$, in $l$, i.e. close to the meridian:
\begin{subequations}
\begin{equation}
\Delta l = \cos\delta\,\,\Delta \alpha.
\end{equation}
\begin{equation}
\therefore \Delta \alpha = \Delta l\,\,\sec\delta.
\end{equation}
\label{e:sinza}
\end{subequations}
\noindent Here, $\Delta \alpha$ is the error in $\alpha$. 
Equation~\ref{e:sinza} shows that an error in $l$ will lead to an error 
in $\alpha$ with a $\sec\delta$-dependence. 

The 2-D images of MRT are 1-D images on the 
meridian made at different sidereal times and stacked. Therefore, positional 
errors both in $\alpha$ and $\delta$ do not show systematics as a function of $\alpha$
({\it first rows} of Figs.~\ref{f:sourcecomparison}a~and~\ref{f:sourcecomparison}b).
We preferred $(\alpha,\sin za)$-representation because all MRT images 
were already generated in this coordinate system. 
This choice compelled us to seek solutions for errors in $\alpha$ as a function of $\sin za$ 
rather than $\sec\delta$.
We plotted errors in $\alpha$ against both $\sec\delta$ and $\sin za$ and obtained 
separate linear least-squares fits. The rms of residuals in both fits 
is $\sim 5\%$ of the beamwidth in $\alpha$. However, the rms of difference between the 
fitting functions $\sec\delta$ and $\sin za$ in the $\delta$ range of MRT
$(-70^{\circ} \mbox{ to } -10^{\circ})$ is only $\sim 1.5\%$ of the beamwidth in $\alpha$. 
Therefore, the random errors in the source positions are larger than the errors 
introduced by the preferred $(\alpha,\sin za)$-coordinates for ${\mathbf p}^{}_{k}$ and ${\mathbf p}'_{k}$.

In ${\mathbf p}^{}_{k}$ and ${\mathbf p}'_{k}$, the $\alpha$ ranges from 18 hours
to 24 hours and the $\sin za$ ranges from $-0.8$ to $0.2$ (corresponding to the declination range
of $-70^{\circ}$ to $-10^{\circ}$). 
Moreover, in matrix $\tt A$ (refer Equation~\ref{e:system}) there are entries of 1's \& 0's.
Such a matrix is ill-conditioned and in the presence of noise in the source positions, the solution 
for an over-determined system may diverge from the correct estimate~\citep{book:hartley}. 
The effect of an ill-conditioned matrix is that it amplifies the divergence. 
A normalisation (or pre-conditioning) is therefore required.

\subsection{Data normalisation and denormalisation}
To obtain a good estimate of the transformation matrix we adopted the normalisation 
scheme proposed by \cite{pami:hartley97}. The normalisation ensures freedom on arbitrary 
choices of scale and coordinate origin, leading to algebraic minimisation in a fixed 
canonical frame. The homography matrix $\tt \tilde{H}$ is estimated from normalised coordinates 
by the least-squares method using singular value decomposition (SVD). The matrix is then 
denormalised to obtain $\tt H$. The scheme is briefly described below:

\begin{enumerate}
\item {\bf Normalisation of $\mathbf p$:} Compute a transformation matrix $\tt M$, consisting
of a translation and scaling, that takes points ${\mathbf p}^{}_{k}$ to a new set of points
$\tilde{\mathbf p}^{}_{k}$ such that the centroid of the points $\tilde{\mathbf p}^{}_{k}$ 
is the coordinate origin $(0,0)^{T}$, and their average distance from the origin is $\sqrt{2}$.

\item {\bf Normalisation of ${\mathbf p}'$:} Compute a similar transformation matrix ${\tt M}'$,
transforming points ${\mathbf p}'_{k}$ to $\tilde{\mathbf p}'_{k}$.

\item {\bf Estimate homography:} Estimate the homography matrix $\tt \tilde{H}$
from the normalised correspondences $\tilde{\mathbf p}^{}_{k} \rightarrow \tilde{\mathbf p}'_{k}$
using the algorithm described earlier in the main section.

\item {\bf Denormalisation:} The final homography matrix is given by: 
\begin{equation}
{\tt H} = {\tt M}'^{-1}\,{\tilde{\tt H}}\,{\tt M}.
\nonumber 
\end{equation}
\end{enumerate}

\begin{figure*}
\centering
\subfigure[]{
\epsfig{figure=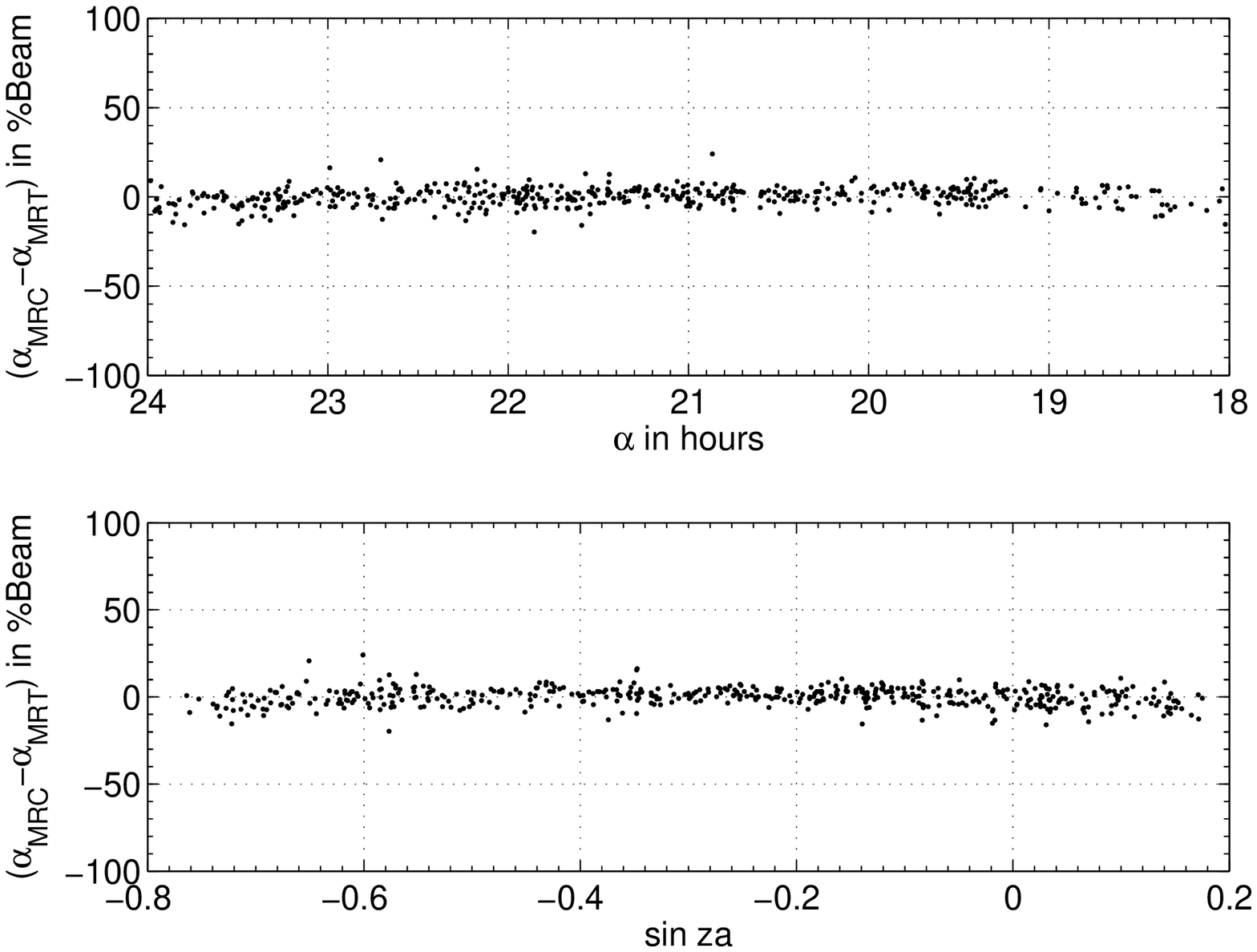,width=0.44\linewidth}}
\subfigure[]{
\epsfig{figure=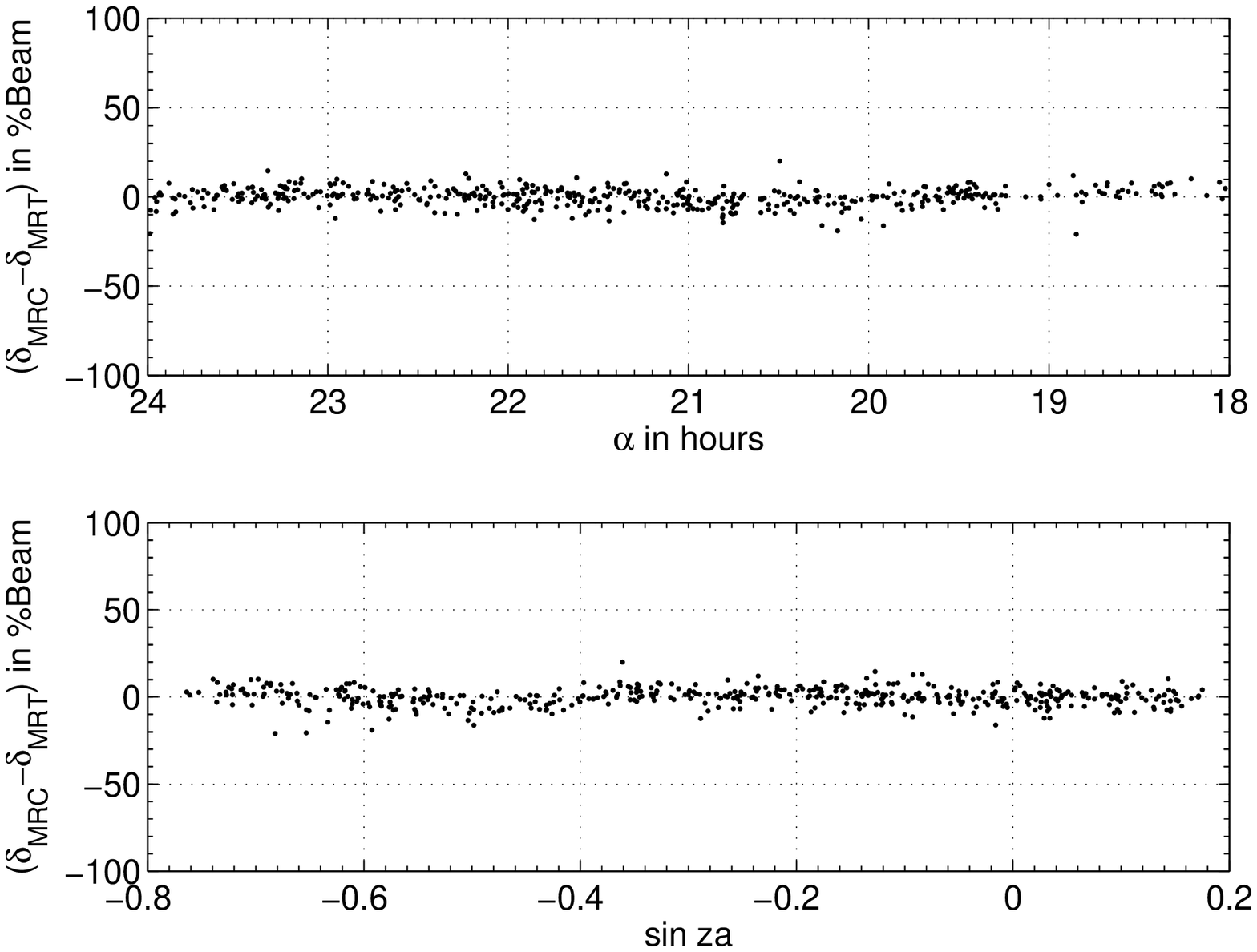,width=0.44\linewidth}}
\subfigure[]{
\epsfig{figure=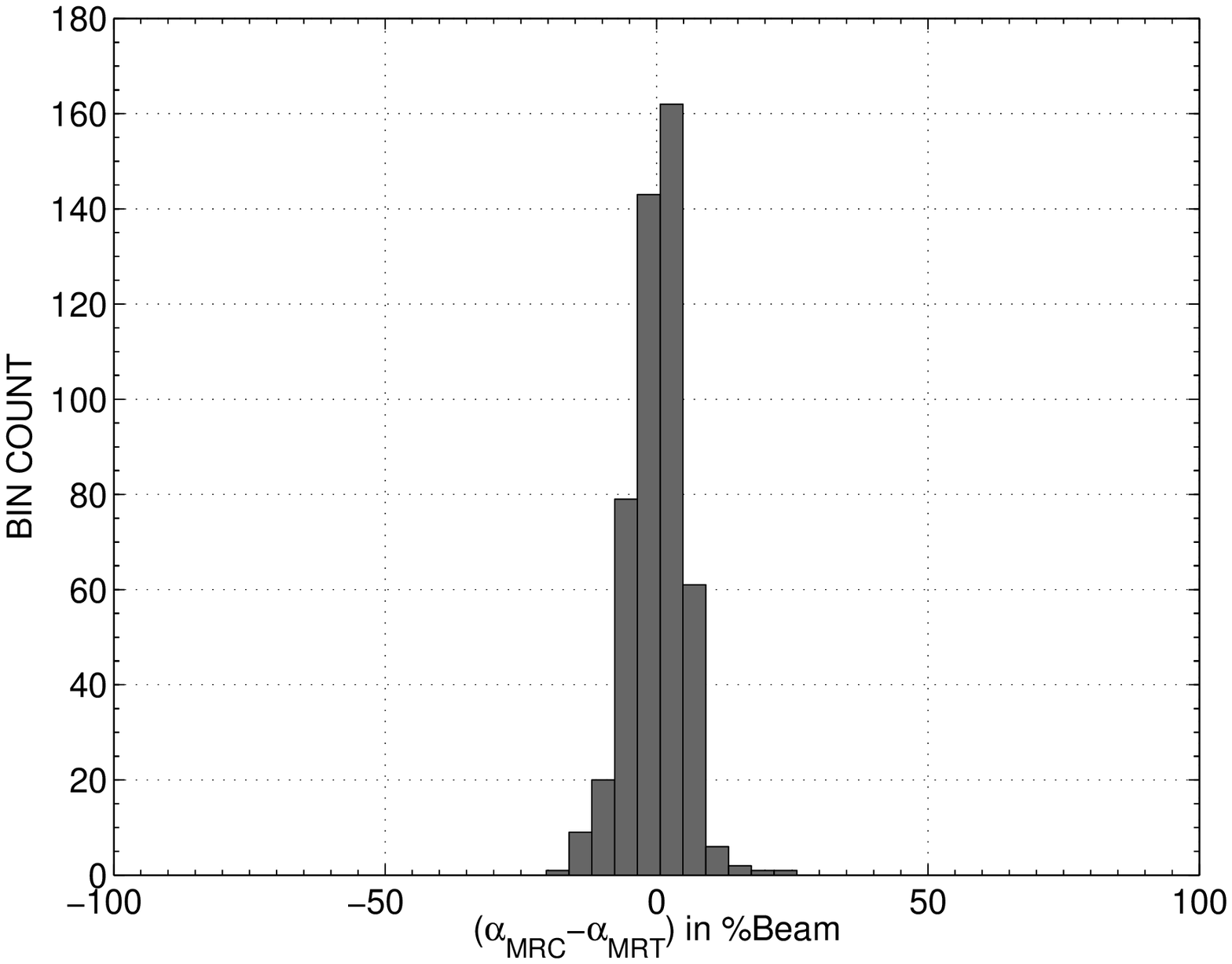,width=0.44\linewidth}}
\subfigure[]{
\epsfig{figure=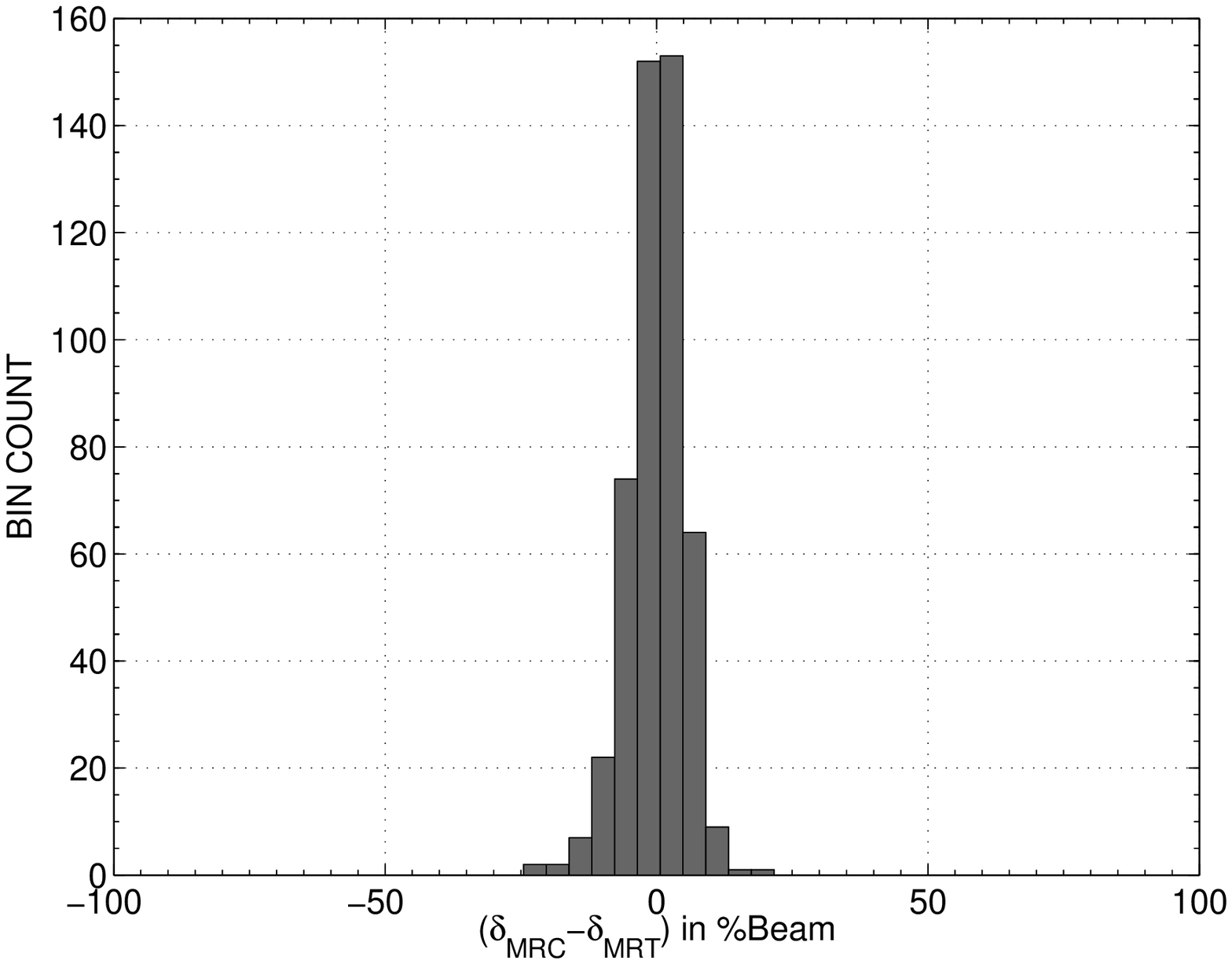,width=0.44\linewidth}}
\caption{\small Positional error analysis after 
homography-based correction. 
{\bf (a)} {\it First} and {\it second row} subplots show errors in $\alpha$ against $\alpha$ and $\sin za$,
respectively. {\bf (b)} The {\it first} and {\it second row} subplots show errors in $\delta $
against $\alpha$ and $\sin za$, respectively. {\bf (c)} and {\bf (d)} show histograms of errors in
$\alpha$ and $\delta$, respectively. 
A comparison of these plots with
Fig.~\ref{f:sourcecomparison} demonstrate that homography has 
removed the systematics and the residual errors are within 10\% of the 
beamwidth.
\normalsize }
\label{f:sourcecomparisoncorrect}
\end{figure*}

\begin{figure*}
\centering
\subfigure[]{
\epsfig{figure=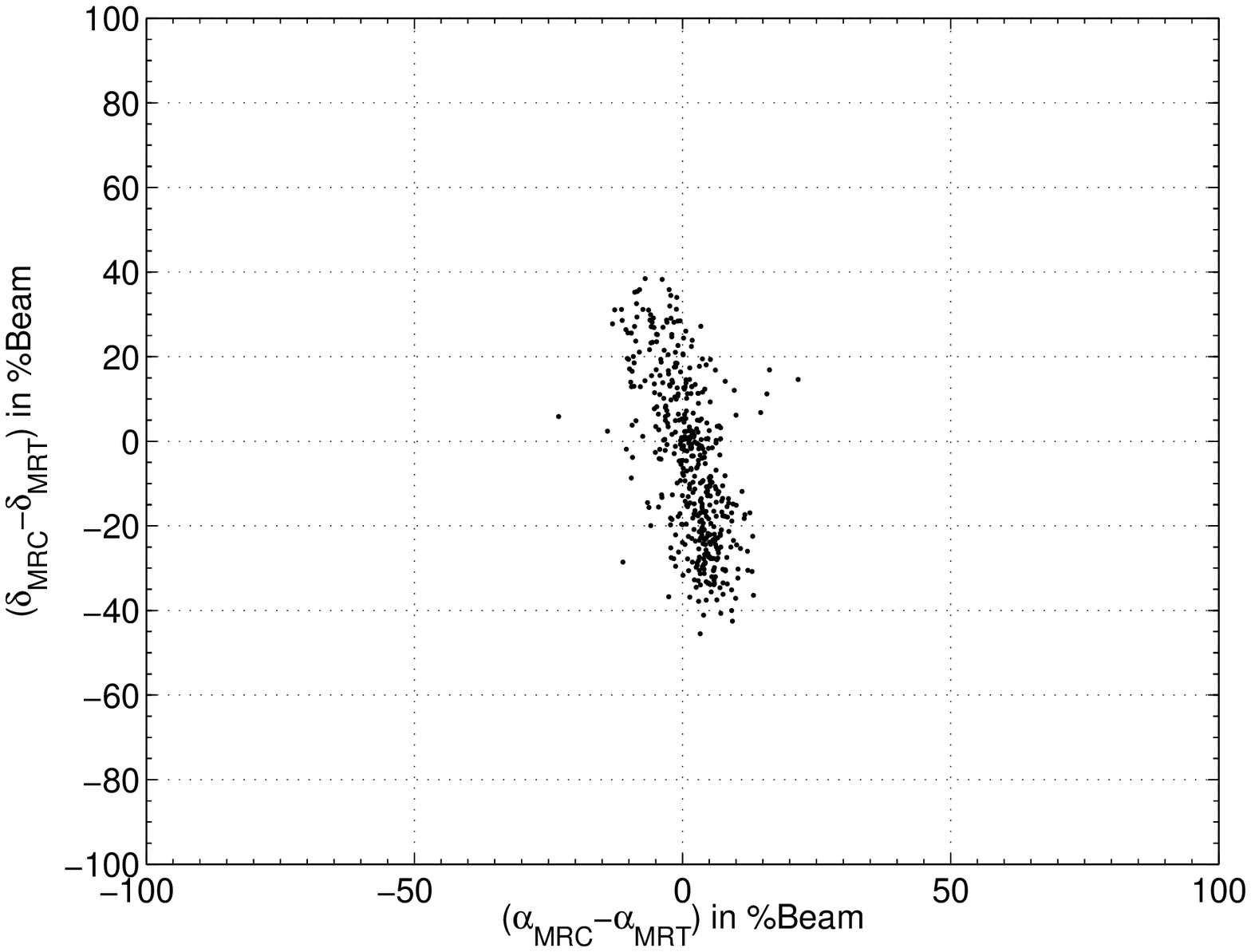,width=0.44\linewidth}}
\subfigure[]{
\epsfig{figure=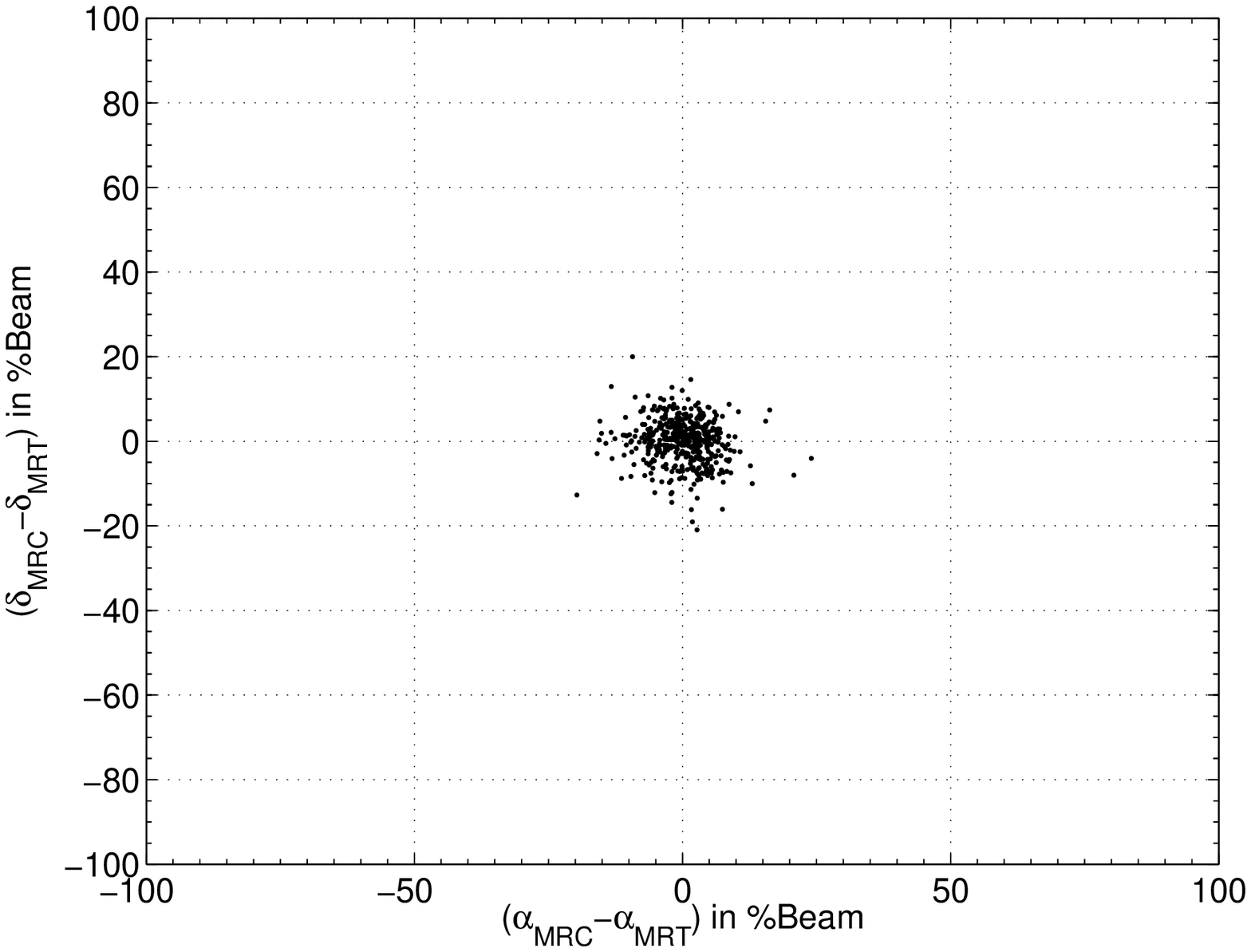,width=0.44\linewidth}}
\caption{\small Scatter plot of errors in $\alpha$ and $\delta$. {\bf (a)} Before correction and {\bf (b)} after homography-based correction.
After correction the scatter is almost circular as opposed to elliptical 
before correction. The rms before correction is $\sim 20\%$ of the beamwidth.
After correction, the rms is reduced to $\sim 7\%$ of the beamwidth and, the 
systematic errors have been removed.}
\label{f:scatterplot}
\end{figure*}

\begin{figure*}
\centering
\subfigure[]{
\epsfig{figure=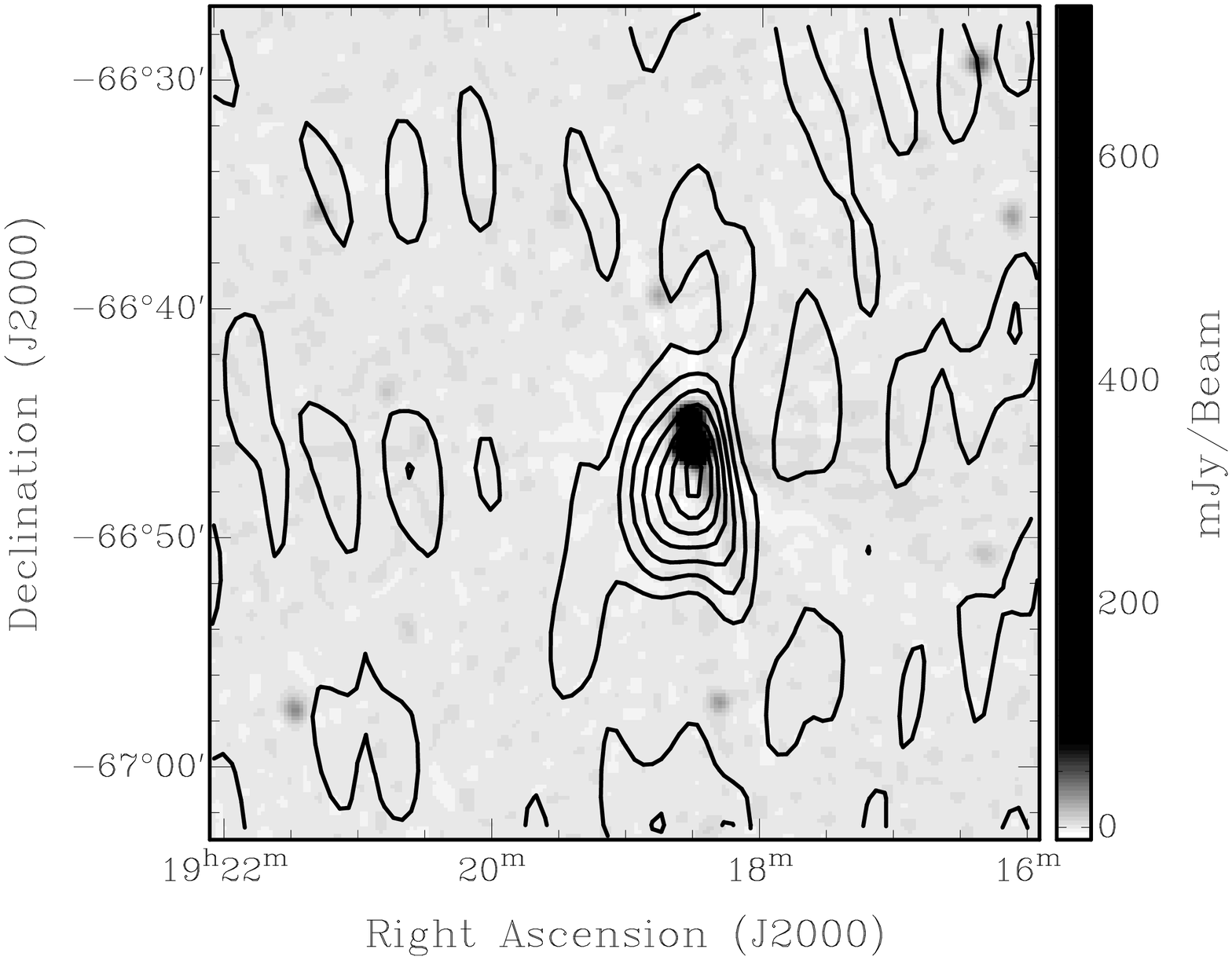,angle=-90,width=0.45\linewidth}}
\subfigure[]{
\epsfig{figure=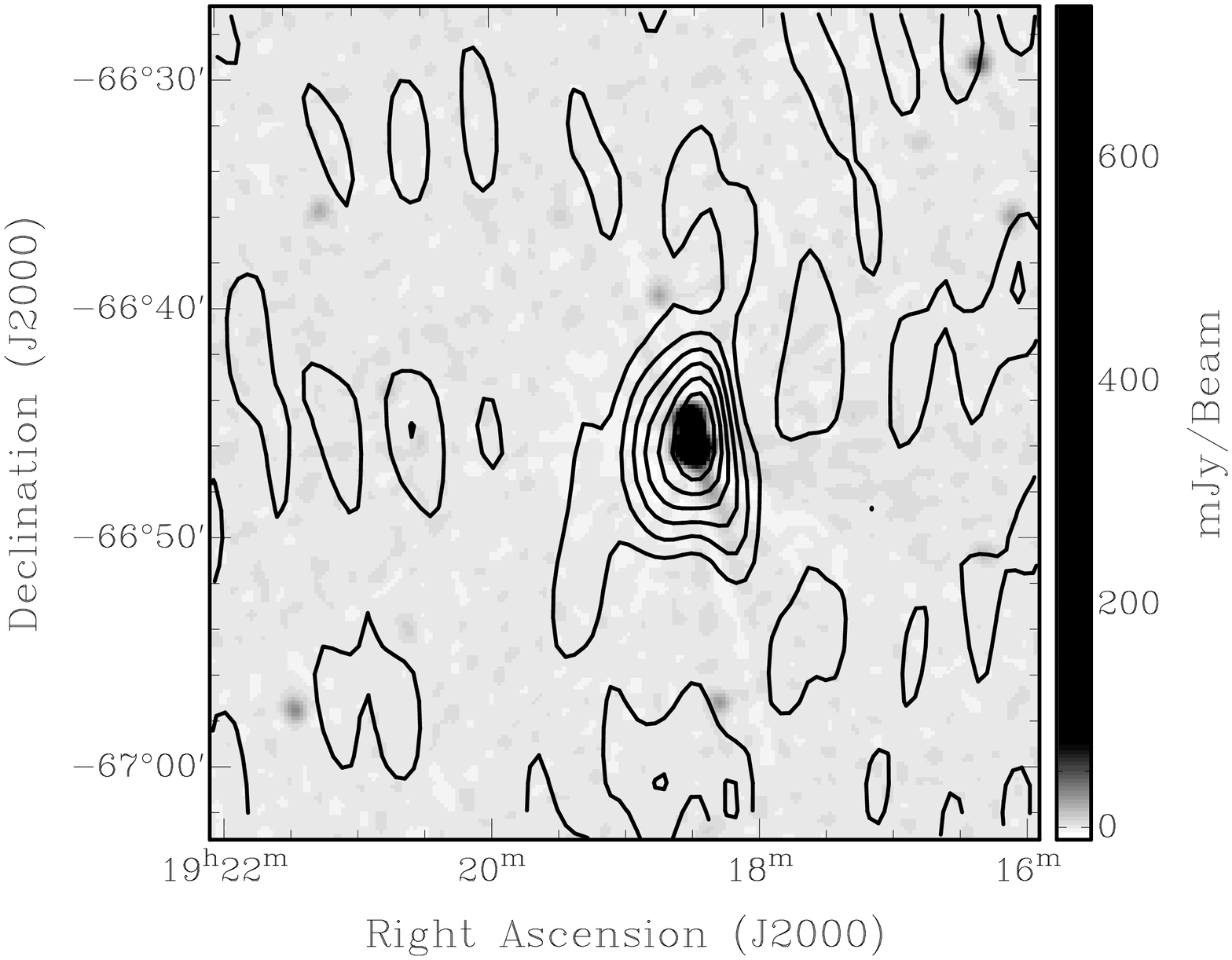,angle=-90,width=0.45\linewidth}}
\subfigure[]{
\epsfig{figure=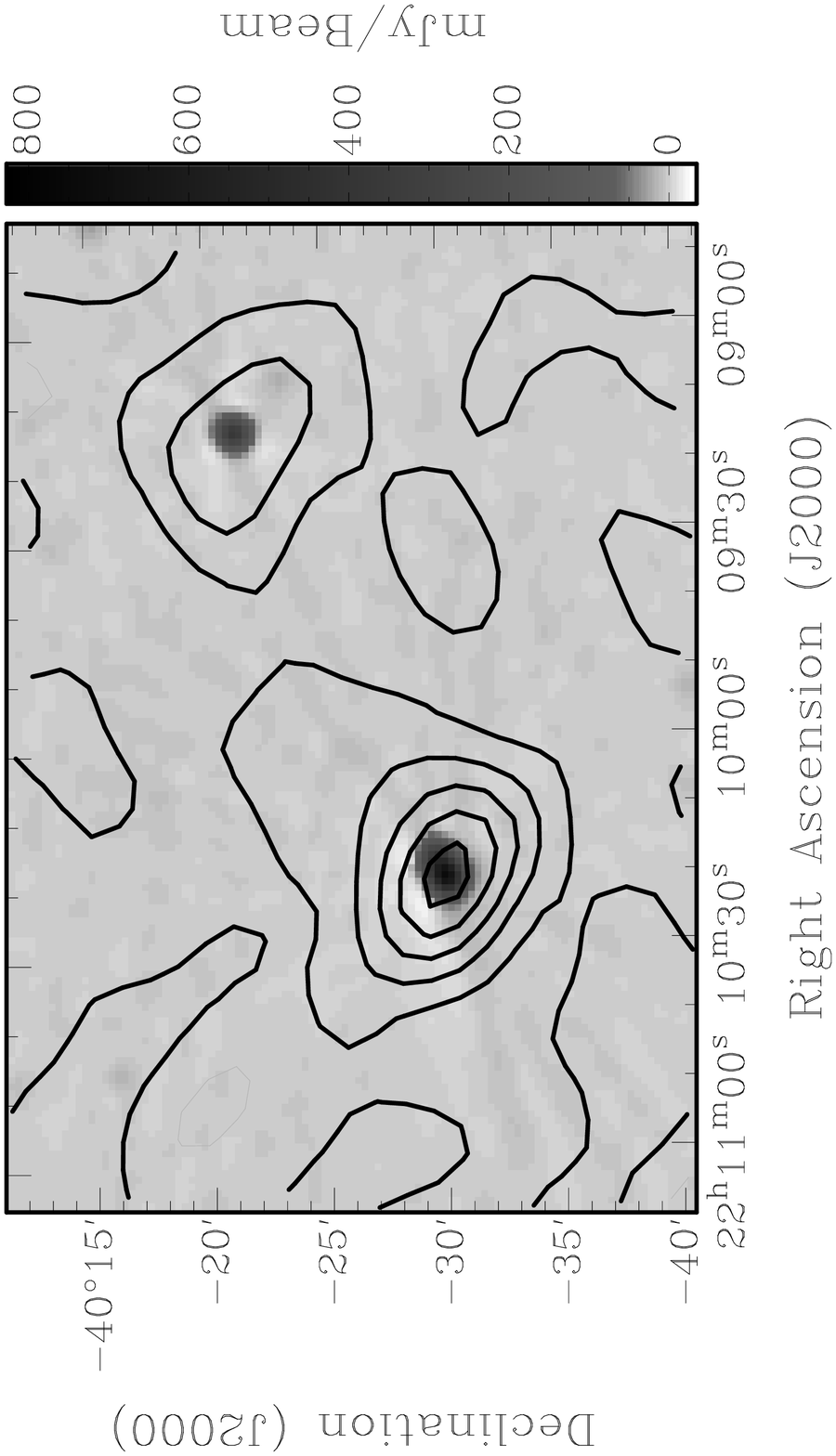,angle=-90,width=0.45\linewidth}}
\subfigure[]{
\epsfig{figure=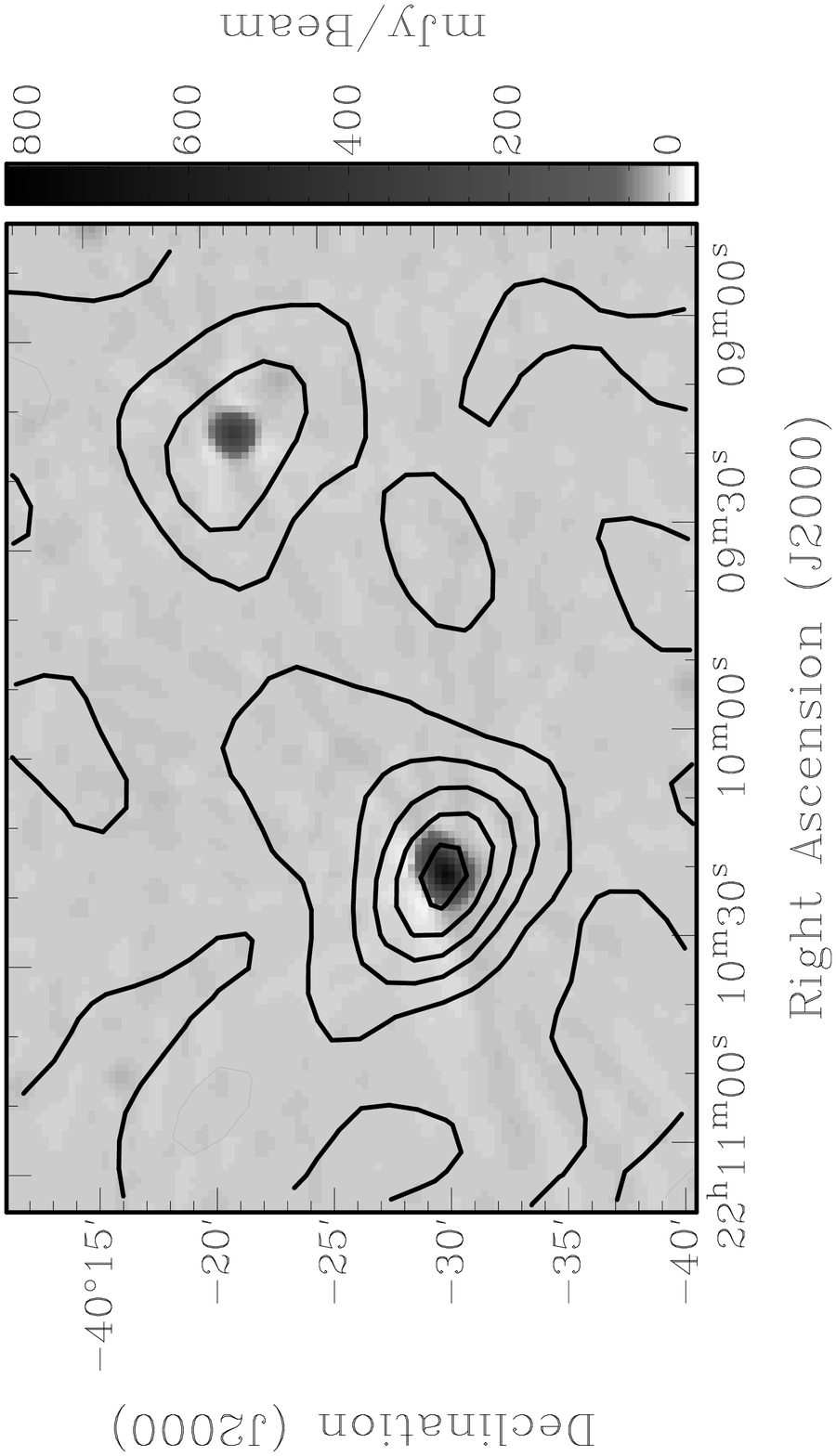,angle=-90,width=0.45\linewidth}}
\caption{ \small MRT contours overlaid on SUMSS image. {\bf (a)} 
and {\bf (c)} MRT contours before correction for sources at about 
$\delta=-66^\circ$ and $\delta=-40^\circ$, respectively. {\bf (b)} and {\bf 
(d)} Corresponding MRT contours after homography-based correction
show that 2-D homography corrected the positional errors.
Notice, (c) and (d) are included here for visual emphasis. Since the errors 
around $\delta=-40^\circ$ are within 10\% of the beamwidth the contours 
show a good overlap both before and after and, as expected homography 
has not applied perceivable correction to images at this declination.
\normalsize}
\label{f:contourplots}
\end{figure*}

\section{The Correction scheme}
\label{s:scheme}

Fig.~\ref{f:scheme} shows the block schematic of the correction scheme.
At MRT, the full declination range for each sidereal hour range is divided into 
4 {\it zones} (refer {\it second row} in 
Fig.~\ref{f:sourcecomparison}a~or~\ref{f:sourcecomparison}b). Each zone 
is imaged with different delay settings to keep the bandwidth 
decorrelation to $< 20\%$. Therefore, the 6 sidereal hours of images under 
consideration, have 24 images ($\sim 15^{\circ}\times15^{\circ}$).

Using the population of common sources, there are four possible alternatives 
to correct MRT images by computing:
\begin{enumerate}
\item 24 homography matrices - one for each image.
\item 6 matrices - one for each sidereal hour.
\item 4 matrices - one for each declination zone.
\item A single homography matrix for the entire steradian.
\end{enumerate}
\noindent In principle, bright sources in each image ($15^{\circ}\times15^{\circ}$) can be used to 
independently estimate a homography matrix. Our earlier experiments to correct 
each image independently showed that the homography matrices were similar.
The plots of errors in $\alpha$ and $\delta$ plotted against $\alpha$ and $\sin za$ (refer to  
Fig.~\ref{f:sourcecomparison}a~and~\ref{f:sourcecomparison}b) indicate 
that the errors are independent of the four delay zones and the range of $\alpha$. 
This implies that estimating a single homography matrix for 
the entire source population should suffice in representing the errors.

The homography matrix estimated using $\sim 400$ common
sources (described in Section~\ref{s:poserror}) is:
\begin{equation}
{\tt H} =
\left[ \begin{array}{ccc}
1.0000 & 0.0006 & \,\,\,\,\,0.0001\\
0.0000 & 0.9990 & -0.0009\\
0.0000 & 0.0000 & \,\,\,\,\,1.0000\\
\end{array}\right].
\label{eq:matrixvalues}
\end{equation}
\noindent In the estimated homography matrix, $h^{}_{11}=1.0000$ 
indicates there is no correction required in $\alpha$ as a 
function of $\alpha$. $h^{}_{12}=0.0006$ indicates MRT images should be corrected
in $\alpha$ with a $\sin za$ dependence. The
estimated correction is up to $\sim 10\%$ of the beam
in $\alpha$, at the extreme ends of the $\sin za$ range.
Similarly, $h^{}_{21}=0.0000$ indicates that there is no correction required
in $\sin za$, as a function of $\alpha$.
However, $h^{}_{22}=0.9990$ indicates that MRT images should be compressed in
$\sin za$ by a factor of 0.9990 (which is $\sim 1$ part in 1000).
The values of $h^{}_{13}$ and $h^{}_{23}$ indicate that the zero cross-overs
of errors in both $\alpha$ and $\sin za$ plotted against $\sin za$
are close to the $\sin za$ of the calibration source (MRC\,1932-464) used for
imaging.

Using Equation~\ref{eq:inhomog}, the homography matrix is used to project each pixel
from the images to a new position, effectively correcting for positional
errors in images.

\subsection{Corrected images and discussion}
\label{ss:results}

\begin{figure}
\centering
\subfigure[]{
\epsfig{figure=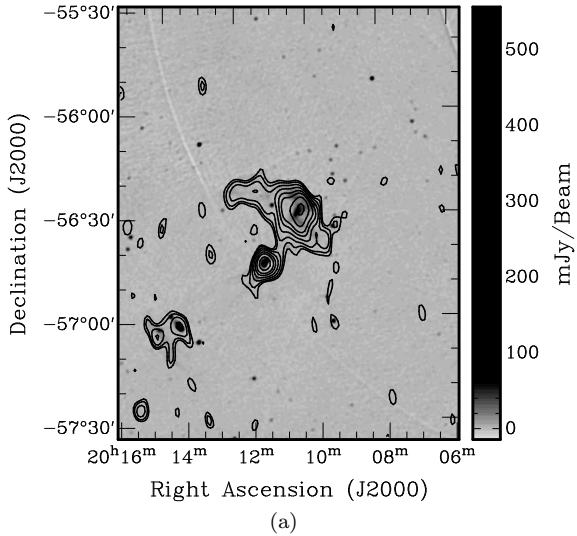,width=0.9\linewidth}}
\caption{ \small MRT contours overlaid on SUMSS image of a region around
Abell 3667. \normalsize}
\label{f:contourplotsextsources}
\end{figure}

Fig.~\ref{f:sourcecomparisoncorrect} shows positional errors in $\delta$ after
homography-based correction. A comparison of these plots with
Fig.~\ref{f:sourcecomparison} demonstrate that homography has 
removed the systematics and the residual errors are within 10\% of the 
beamwidth for sources above 15-$\sigma$, as expected. 
Fig.~\ref{f:scatterplot}a and Fig.~\ref{f:scatterplot}b show scatter plots 
of errors in $\delta$ against errors in $\alpha$ before and after correction, 
respectively. 
For visualisation, the errors are represented in percentages of respective 
MRT beamwidths.
Notice, after correction (refer Fig.~\ref{f:scatterplot}b) the scatter is
almost circular as opposed to elliptical before correction (refer Fig.~\ref{f:scatterplot}a).
The rms before correction is $\sim 20\%$ of the beamwidth. After correction, 
the rms is reduced to $\sim 7\%$ of the beamwidth and, the systematic errors 
have been removed.

Fig.~\ref{f:contourplots}a~and~\ref{f:contourplots}b 
show MRT contours before and after correction, respectively, overlaid 
on SUMSS (Sydney University Molonglo Sky Survey) image
\citep{mnras:mauch03}, for a source around $\delta = -67^\circ$. The corrected
MRT image contours in Fig.~\ref{f:contourplots}b overlap with the
source in SUMSS image.
Figs.~\ref{f:contourplots}c~and~\ref{f:contourplots}d show similar
comparison for a source around $\delta=-40^\circ$. Notice
Fig.~\ref{f:sourcecomparison}d, since the errors around $\delta=-40^\circ$
are within 10\% of the beamwidth, the contours in
both Figs.~\ref{f:contourplots}c~and~\ref{f:contourplots}d show a good
overlap as expected and homography has not applied perceivable correction
to images at this declination.
We have overlaid MRT contours on a number of extended sources at 843~MHz
reported by \cite{apjss:jones92}. Fig.~\ref{f:contourplotsextsources} shows
a typical overlay of MRT contours on SUMSS image of a region 
around the cluster Abell 3667. The overlay is perceivably satisfactory.

The 2-D homography corrected the positional errors in the image domain.
For imaging the remaining $\sim 3.5$ steradians of MRT survey, $\sim 15000$ hours 
of data has to be reduced. Ideally, 
for imaging the new regions, one would like to trace the source of these errors and 
correct them in the visibilities. In the following section we discuss how we traced 
the source of errors and corrected them in the visibility domain.

\section{Array Geometry: Hypothesis \& Re-estimation}
\label{s:arraygeometry}
This section describes our {\it expansion-compression}
hypothesis for the source of errors in our images. The
subsequent corrections we estimated and applied to 
eliminate the errors are also described.

For meridian transit imaging, $m = \sin za$.
The brightness distribution in the sky as a function of $\sin za$ and
the complex visibilities measured for different values of the north-south (NS) baseline vector
component $v$ form a Fourier pair~\citep{book:christiansen}.
A scaling error of $\kappa$ in $m$ will result in a scaling factor of
$\kappa^{-1}$ in the $v$-component of the baseline vector. By positional error 
analysis it is clear that MRT images are stretched ({\it expanded}) in declination, i.e.,
\begin{subequations}
\begin{equation}
m^{}_{\mbox{\small imaged}} = \kappa\,m^{}_{{\mbox{\small true}}}
\end{equation}
\begin{equation}
\therefore v^{}_{\mbox{\small measured}} = \kappa^{-1}\,v^{}_{\mbox{\small true}}.
\end{equation}
\label{e:expcomp}
\end{subequations}
\noindent Note, for images the 2-D homography estimated a correction 
({\it compression}) factor, $\kappa^{-1}$, of 0.9990. 
This cued to the hypothesis that we have compressed the north-south
baseline vectors.
Equation~\ref{e:expcomp}b means, a baseline distance of 
$\sim 1000$\,m in the NS arm was wrongly measured as $\sim 999$\,m (1 part in 1000). 
Similarly, a $\sin za$-dependent correction in $\alpha$ cued to
possible $v$-component in the east-west (EW) baseline vectors.
Next, we describe the re-estimation of array geometry.

We begin with a brief description of the mode of observations with MRT.
MRT has 32 fixed antennas in the EW arm and 15 movable antenna trolleys in the 
NS arm. For measuring visibilities, the 15 NS trolleys are configured by spreading them 
over 84~m with an inter-trolley 
spacing of 6~m (to avoid shadowing of one trolley by another). MRT measures different 
Fourier components of the brightness distribution of the sky in 63 different 
configurations (referred to as {\it allocations}) to sample NS baselines every 1\,m.
Therefore, effectively, there are 945 antenna positions (63 allocations * 15 antennas/allocation)
in the NS arm and a total of 30,240 (945 * 32) visibilities are used for imaging.

\begin{figure}
\centering
\subfigure[]{
\epsfig{figure=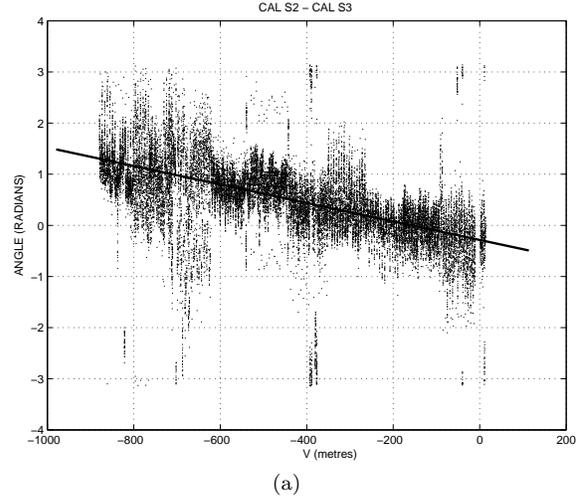,width=0.9\linewidth}}
\caption{\small Typical calibrator phase differences (in radians)
of MRC\,0915-118 \& MRC\,1932-464, 
plotted against $v$ (in metres). The straight line shown is
a linear robust fit obtained for the data.\normalsize}
\label{f:phasediff}
\end{figure}

\begin{figure*}
\centering
\subfigure[]{
\epsfig{figure=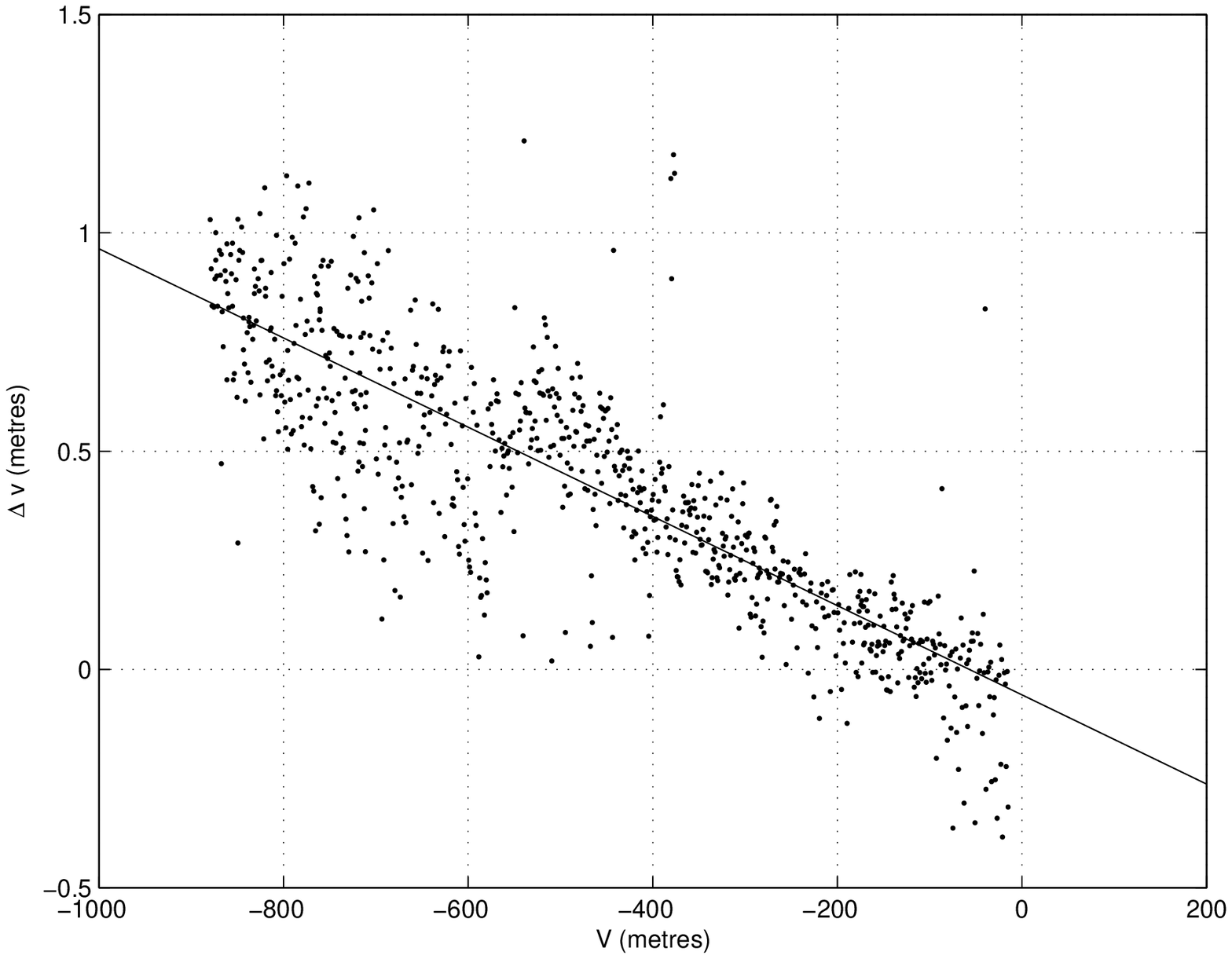,width=0.45\linewidth}}
\subfigure[]{
\epsfig{figure=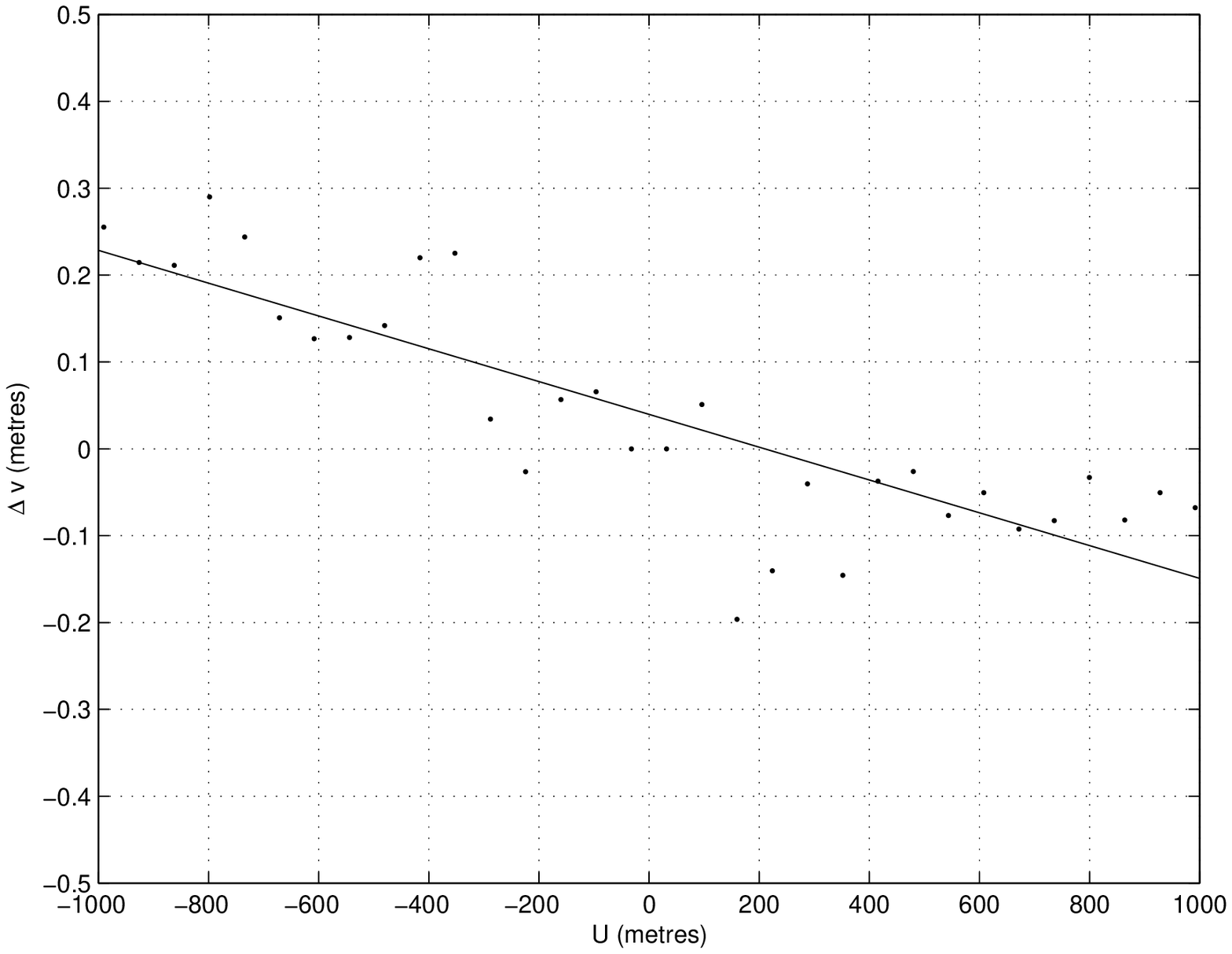,width=0.45\linewidth}}
\caption{\small {\bf (a)} Estimate of the antenna position along the North-South arm
of MRT array. The fit to the estimates shows a gradient of 1 part in 1000, along
the north-south.
{\bf (b)} Estimate of the error in $v$ coordinate of MRT east-west arm. The fit to the
estimate shows a gradient of 2 part in 10,000, along the east-west.}
\label{f:error_estimate}
\end{figure*}

A small error in a measuring scale of relatively shorter length is likely to build up 
systematically while establishing the geometry of longer baselines. This effect would 
be observed in the instrumental phases estimated using different calibrators.
In principle, the instrumental phases estimated using two calibrators
at different declinations, for a given baseline, 
should be the same, allowing for temporal variations in the instrumental gains.
A non-zero difference in these estimates may be due to positional
errors of the baseline or positions of calibrators. As mentioned earlier,
our analysis of positional error in sources and the homography matrix
cued to positional errors in baselines (or antenna positions).
The simple principle of astrometry~\citep{book:thomson} was used
to estimate errors in antenna positions and is discussed below.

The observed visibility phase, $\psi_{ij}^{\mathcal{S}^{}_1}$, in a baseline with components 
$\left(u^{}_{ij}, v^{}_{ij}, w^{}_{ij}\right)$, due to calibrator $S^{}_{1}$ with direction
cosines $\left(l^{\mathcal{S}_1},m^{\mathcal{S}_1},n^{\mathcal{S}_1}\right)$, is given by:
\begin{equation}
\psi_{ij}^{\mathcal{S}^{}_{1}} = l^{\mathcal{S}^{}_{1}} u^{}_{ij} + m^{\mathcal{S}^{}_{1}} v^{}_{ij} + n^{\mathcal{S}^{}_{1}} w^{}_{ij} + \phi_{ij}^{\mbox{\small ins}}.
\label{e:obsphasebasiceqn}
\end{equation}
\noindent Where, $\phi_{ij}^{\mbox{\small ins}}$ represents true instrumental phases, $i=1,2,\ldots,32$ represents
EW antennas and $j=1,2,\ldots,945$ represents NS antennas. For meridian transit imaging Equation~\ref{e:obsphasebasiceqn} becomes: 
\begin{equation}
\psi_{ij}^{\mathcal{S}^{}_{1}} = - v^{}_{ij} \sin\left(ZA^{\mathcal{S}^{}_{1}}\right) + w^{}_{ij} \cos\left(ZA^{\mathcal{S}_1}\right) +
\phi_{ij}^{\mbox{\small ins}}.
\end{equation}
The instrumental phases, $\phi_{ij}^{\mathcal{S}^{}_{1}}$,
estimated using the measured geometry are given by:
\begin{equation}
\phi_{ij}^{\mathcal{S}^{}_{1}} = - \Delta v^{}_{ij} \sin\left(ZA^{\mathcal{S}^{}_{1}}\right) + 
\Delta w^{}_{ij} \cos\left(ZA^{\mathcal{S}^{}_{1}}\right) + \phi_{ij}^{\mbox{\small ins}}.
\label{eq:calphaserelation}
\end{equation}
\noindent Here, $\Delta v^{}_{ij}$ and $\Delta w^{}_{ij}$ are errors in the assumed
baseline vectors. $\phi_{ij}^{\mathcal{S}^{}_{1}}$ are phases of complex
baseline gains obtained in the process of calibration.
\noindent Equation~\ref{eq:calphaserelation} has three unknowns.
To reduce the number of unknowns, one can eliminate the true instrumental
phases by taking a difference $\left(\Delta\phi_{ij}^{\mathcal{S}^{}_{12}} = \phi_{ij}^{\mathcal{S}^{}_{1}} - \phi_{ij}^{\mathcal{S}^{}_{2}}\right)$ between the instrumental phases estimated using two calibrators.
This difference gives:
\begin{eqnarray}
\lefteqn{\hspace{10mm}\Delta\phi_{ij}^{\mathcal{S}^{}_{12}} = - \Delta v^{}_{ij} \left[ \sin\left(ZA^{\mathcal{S}^{}_{1}}\right) - \sin\left(ZA^{\mathcal{S}^{}_{2}}\right)\right]} \nonumber \\
\lefteqn{\hspace{22mm} + \Delta w^{}_{ij} \left[\cos\left(ZA^{\mathcal{S}^{}_{1}}\right) - \cos\left(ZA^{\mathcal{S}^{}_{2}}\right)\right].}
\label{eq:diff1}
\end{eqnarray}
\noindent 
Note, the $w$-components of the baseline vectors are short and non-cumulative measurements.
Therefore, in principle, one can consider $\Delta w^{}_{ij}$ as zero-mean random 
errors with no systematics. Equation~\ref{eq:diff1} in that case can be written as:
\begin{equation}
\Delta\phi_{ij}^{\mathcal{S}^{}_{12}} = - \Delta v^{}_{ij} \left[\sin\left(ZA^{\mathcal{S}^{}_{1}}\right) - \sin\left(ZA^{\mathcal{S}^{}_{2}}\right)\right].
\label{eq:simplediff1}
\end{equation}
\noindent Describing the system in terms of errors in antenna 
positions, as opposed to errors in baseline positions, Equation~\ref{eq:simplediff1} becomes:
\begin{equation}
\Delta\phi_{ij}^{\mathcal{S}^{}_{12}} = - \left(\Delta v^{}_{i}-\Delta v^{}_{j}\right) \left[\sin\left(ZA^{\mathcal{S}^{}_{1}}\right) - \sin\left(ZA^{\mathcal{S}^{}_{2}}\right)\right].
\label{eq:diffantenna1}
\end{equation}
\noindent This equation is also not sufficient to solve for errors in 
the antenna positions as we have two unknowns and one equation. We set up 
another equation using a third calibrator source, $\mathcal{S}^{}_{3}$, spaced away in 
declination from $\mathcal{S}^{}_{1}$ and $\mathcal{S}^{}_{2}$:
\begin{equation}
\Delta\phi_{ij}^{\mathcal{S}^{}_{23}} = - \left(\Delta v^{}_{i}-\Delta v^{}_{j}\right) \left[\sin\left(ZA^{\mathcal{S}^{}_{2}}\right) - \sin\left(ZA^{\mathcal{S}^{}_{3}}\right)\right].
\label{eq:diffantenna2}
\end{equation} 
\noindent The Equations~\ref{eq:diffantenna1}~and~\ref{eq:diffantenna2} are a linear 
set of equations for one baseline. For the measurements in 63 allocations, the set of 
equations can be formulated in a matrix form and solved by SVD-based least-squares estimator:
\begin{equation}
{\mathcal A}{\mathbf x}={\mathbf b}.
\end{equation}
Where, the {\it measurement vector} ${\mathbf x} \in \mathbb{R}^{c}$ is to be determined.
Here, $c=977$. The measurement vector gives $\Delta v^{}_{i}$ and $\Delta v^{}_{j}$ estimates
for 32 EW and 945 NS antenna locations, respectively.
The {\it observation vector} $\mathbf b$ consists of two sub-matrices, ${\mathbf b^{}_{1}}\in\mathbb{R}^{r^{}_{1}}$ and 
${\mathbf b^{}_{2}}\in\mathbb{R}^{r^{}_{2}}$, formed using the left-hand-side of Equations~\ref{eq:diffantenna1}~and~\ref{eq:diffantenna2}, respectively.
Here, $r^{}_{1} = r^{}_{2} = 30240$, i.e., the total number of visibilities measured for imaging. 
Therefore, ${\mathbf b} \in \mathbb{R}^{60480}$.
The {\it data matrix} ${\mathcal A} \in \mathbb{R}^{60480\times 977}$.
Each row in the data matrix has only two non-zero elements, corresponding to a baseline 
formed by one EW and one NS antenna, making it very sparse.

The observation vector is constructed from the gain tables of the array 
obtained using calibrators MRC\,0407-658 (${\mathcal{S}^{}_{1}}$), 
MRC\,0915-118 (${\mathcal{S}^{}_{2}}$) and MRC\,1932-464 (${\mathcal{S}^{}_{3}}$). 
The sensitivity per baseline at MRT is $\sim 26$~Jy for a 1~MHz bandwidth and
an integration time of one second. It takes $\sim 10$ minutes of time for sources
at $\delta = -40^\circ$ to transit a 2$^\circ$ primary beamwidth of
elements in the east-west array. This leads to a sensitivity per baseline 
(including the non-uniform weighting due to primary beam) of $\sim 2$~Jy. 
The flux density of these three calibrators as seen by MRT is $\sim 100$~Jy; 
strong to get reliable calibration. Further, the calibrators are unresolved
and isolated from confusing sources and have well known measured 
positions~\citep{thesis:golap98}.

A plot of typical phase differences obtained using the pair of calibrators 
${\mathcal{S}^{}_{2}}\,\,\mbox{and}\,\,{\mathcal{S}^{}_{3}}$
is shown in Fig.~\ref{f:phasediff}.
Fig.~\ref{f:error_estimate}a shows the estimated errors in 945 NS 
antenna positions. The errors show a gradient of 1 part in 1000 along the 
NS arm. This matches with the linear gradients in the phase differences 
estimated from the calibrators.
The estimates in Fig.~\ref{f:error_estimate}b show alignment 
errors of the 32 antennas in the EW arm along the NS-direction. 
The fit shows a gradient of about 2 part in 10,000.
This indicates that the EW arm is mis-aligned from the true EW-direction.
At one extreme end (1 km from the centre of the array) of the EW arm the error is $\sim 0.2$~m, equivalent
to an angular distance of $\sim 40''$ from the centre of the array.
This is the source of a small $\sin za$-dependent error in $\alpha$ that was observed 
in both positional error analysis and the homography matrix. Further, our
simulation of the synthesised beam in $\alpha$ with old EW antenna positions and
the corrected EW antenna positions indeed confirm this $\sin za$-dependent error in $\alpha$.

Using the new antenna positions we have re-imaged one hour from the steradian
and have also imaged a completely new steradian. We find no systematics in 
positional errors thus endorsing our re-estimated array geometry.

\section{Conclusions}
\label{s:conclusions}

The homography-based correction was able to correct for systematics 
in positional errors in the image domain and the errors are 
within 10\% of the beamwidth for sources above 15-$\sigma$.
The corrected images of one steradian are available for download at 
{\it http://www.rri.res.in/surveys/MRT}.

Positional error analysis showed that uncorrected MRT images are 
stretched in declination by $\sim 1$ part in 1000. This translates to a 
compression of the NS baseline vector, in the visibility domain.  The analysis
also showed a $\sin za$-dependent error in $\alpha$. This cued towards possible 
errors in our estimation of the array geometry. By formulating a 
linear system, using instrumental phases estimated from three well 
separated calibrators whose positions are well known, the array geometry 
was re-estimated. The estimated error in the $v$-component of the NS baseline 
vectors is about 1~mm/m. In other words, the error 
is about half a wavelength at 150~MHz (1~m) for a 1~km baseline. 
The estimates also show a small (2 part in 10,000) $v$-component in
the purely EW baseline vectors.
This indicates that the EW arm is mis-aligned and inclined at an angle of
$\sim 40\arcsec$, to the true EW direction.
These estimates match with the observed stretching of MRT images shown by
both the positional error analysis and the homography matrix.

Using the new antenna positions we have re-imaged one hour from the steradian
and have also imaged a completely new steradian. We find no systematics in 
positional errors. This endorses our re-estimated array geometry.
Re-imaging one steradian starting from visibilities would have been
a very time consuming exercise. Development of 2-D homography-based
correction enabled us to correct for the positional errors in the
image domain. In our view, this new technique will be
of relevance to the new generation radio telescopes where, owing to huge
data rates, only images after a certain integration would be
recorded as opposed to raw visibilities.

\section*{Acknowledgement}
Soobash Daiboo acknowledges a PhD bursary from the South African 
Square Kilometer Array project. The authors would like to thank the
anonymous referee for the constructive comments and suggestions.



\label{lastpage}
\end{document}